\documentclass[12pt]{article}\pdfoutput=1
\usepackage{graphics, color}
\usepackage{graphicx}
\usepackage{amsmath}
\usepackage{amssymb}
\usepackage{amsfonts}

\newcommand{\sect}[1]{\section{#1}\setcounter{equation}{0}}

\def\gsim{\, \rlap{$>$}{\lower 1.1ex\hbox{$\sim$}}\,}
\def\lsim{\, \rlap{$<$}{\lower 1.1ex\hbox{$\sim$}}\,}

\newcommand{\be}{\begin{equation}}
\newcommand{\ee}{\end{equation}}
\newcommand{\bea}{\begin{eqnarray}}
\newcommand{\eea}{\end{eqnarray}}

\begin{document}


\begin{titlepage}
\bigskip
\bigskip\bigskip\bigskip
\centerline{\Large Chaos in the black hole S-matrix}
\bigskip\bigskip\bigskip
\bigskip\bigskip\bigskip

 \centerline{ {\bf Joseph Polchinski}\footnote{\tt joep@kitp.ucsb.edu}}
\medskip
\centerline{\em Kavli Institute for Theoretical Physics}
\centerline{\em University of California}
\centerline{\em Santa Barbara, CA 93106-4030}\bigskip
\bigskip
\bigskip\bigskip


\begin{abstract}
Recent work by Shenker, Stanford, and Kitaev has related the black hole horizon geometry to chaotic behavior.  We extend this from eternal black holes to black holes that form and then evaporate.  This leads to an identity for the change in the black hole S-matrix (over times shorter than the scrambling time) due an addition infalling particle, elaborating an idea of 't Hooft.  
\end{abstract}
\end{titlepage}
\baselineskip = 17pt
\setcounter{footnote}{0}


\sect{In what sense is a black hole chaotic?}

Thermal behavior and chaos are intimately connected.  In thermalizing systems, the ergodic mixing of the phase space arises from the exponential divergence of nearby trajectories, Fig. 1ab.  It has been recognized for more than four decades that black holes have thermodynamic properties.  
\begin{figure}[!b]
\begin{center}
\vspace {-5pt}
\includegraphics[width=5.5 in]{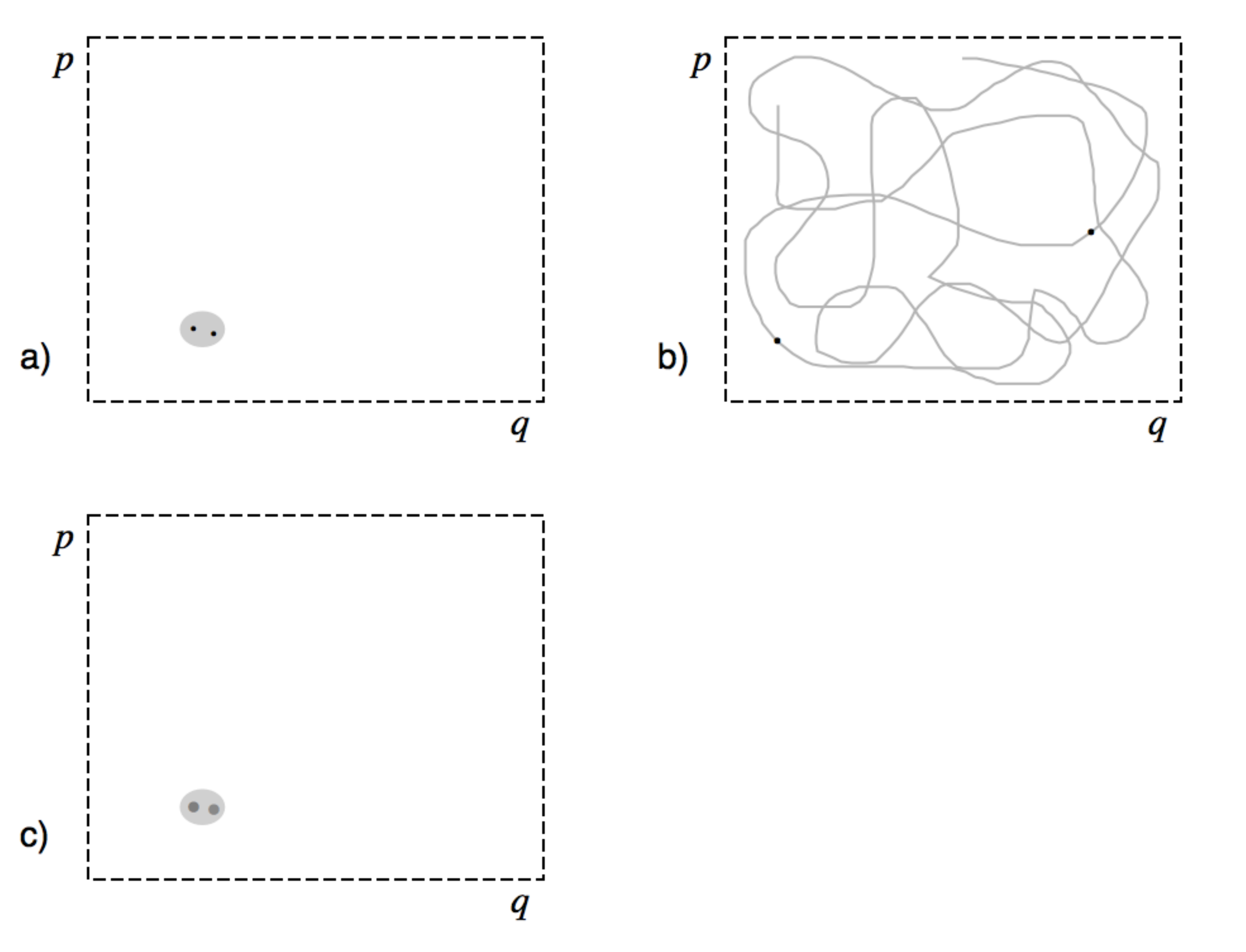}
\end{center}
\vspace {-10pt}
\caption{Phase space of a thermalizing system.  a) Initial classical phase space region.  b) A few thermalization times later, the region is well-mixed into the full phase space (subject to conservation laws).  The initial marked points are now well separated.  c) At small $\hbar$, the initial Wigner distributions still have little overlap.
}
\label{fig:phase}
\end{figure}
However, only very recently has the connection with chaos been made~\cite{Shenker:2013pqa,Shenker:2013yza,Leichenauer:2014nxa,kitaev,Shenker:2014cwa,Jackson:2014nla,Maldacena:2015waa}.  These papers have focused on eternal black holes confined to cavities, either single black holes or in most cases pairs connected by an Einstein-Rosen bridge.  In this paper we extend these ideas to black holes that form and then decay.

A measure of chaos is the sensitivity to initial conditions,
\be
\frac{\partial q(t)}{\partial q(0)} = \{ q(t) , p(0) \}  \,.
\ee
The corresponding quantum quantity would be
\be
\frac{1}{i\hbar} [ \hat q(t) , \hat p(0) ] \,.
\ee
Of course, quantum mechanics is linear, and so two states that are close together in the sense of having a large inner product will remain so.  But states that are orthogonal can still be physically similar (Fig.~1c), and this will not be preserved by time evolution.  A useful quantity is the square of the commutator (to avoid phase cancellations) averaged over the thermal ensemble~\cite{LO},
\be
Z^{-1} {\rm Tr} \left( e^{-\beta H} [ \hat W(t) , \hat V(0) ] [ \hat W(t) , \hat V(0) ]^\dagger \right) \,. \label{comm2}
\ee
We have generalized the operators, conforming to common notation.

A typical behavior for this expectation value is shown in Fig.~2.  
\begin{figure}[!b]
\begin{center}
\vspace {-5pt}
\includegraphics[width=4.5 in]{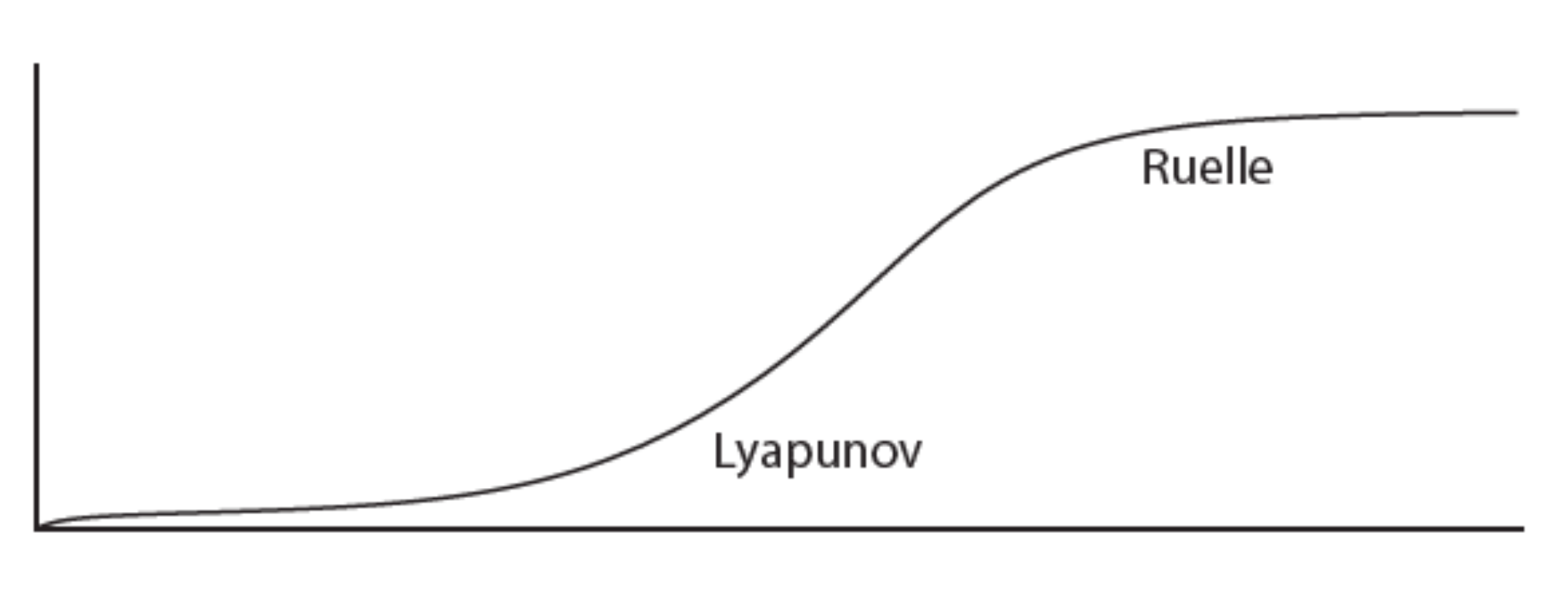}
\end{center}
\vspace {-10pt}
\caption{Commutator-squared versus time.  The exponential Lyapunov growth and Ruelle decay are noted.
}
\label{fig:c2}
\end{figure}
We suppose that the commutator is initially zero (as for spins separated in space, for example).  There are two exponentials of interest.  The first is the exponential decay in the approach to equilibrium, whose exponents are the Ruelle resonances.  The second is the early exponential growth, whose exponents are the Lyapunov exponents (times two, since we have squared the commutator).  Note that the approach to equilibrium can be seen even in two-point functions, but to see the Lyapunov behavior requires the commutator-squared.

In a black hole, there are two notable exponential behaviors.  If a source of fixed frequency is thrown into a black hole, the frequency seen at infinity will be exponentially reshifted,
\be
\frac{d\tau}{dt} \propto e^{-2\pi t/\beta} \,. \label{redshift}
\ee
Here $\tau$ is the proper time of the infalling source, $t$ is the Schwarzschild time, and $\beta$ is the inverse Hawking temperature.  The other exponential is seen in the decay of a perturbed black hole toward its hairless state.  For a Schwarzschild black hole the quasinormal modes governing this decay have time constants of the same order as in the redshift~(\ref{redshift}), since there is only one time scale in the system, but they are not the same.  Rather, these modes are found by solving a differential equation involving distances of order $r_{\rm s}$ outside the black hole, whereas the redshift~(\ref{redshift}) depends only on the temperature which is determined entirely by the geometry at the horizon (the surface gravity).

Clearly the quasinormal modes correspond to the Ruelle resonances.  The recent observation, as we will explain further below, is that the redshift~(\ref{redshift}) is the sign of Lyapunov growth~\cite{kitaev,Shenker:2014cwa,Maldacena:2015waa}.  The asymptotic observer actually detects the reciprocal, ${dt}/{d\tau}$, which is a growing exponential.  It has long been of interest to understand how the black hole horizon is manifested in the dual quantum theory.  A general connection with thermalization was obvious, but the work of~\cite{Shenker:2013pqa,Shenker:2013yza,Leichenauer:2014nxa,kitaev,Shenker:2014cwa,Jackson:2014nla,Maldacena:2015waa} now makes it clear that it is the Lyapunov behavior that is a direct reflection of the horizon physics.\footnote{There is some similarity between the recently proposed model of horizon chaos~\cite{kitaev} and earlier matrix models of black hole thermalization~\cite{Festuccia:2006sa,Iizuka:2008hg,Iizuka:2008eb} (see in particular Figs.~1 of~\cite{Festuccia:2006sa,Iizuka:2008hg}).  However, it is not clear that any of the latter share the property of being both solvable and chaotic.  I thank B. Michel, V. Rosenhaus, and S. J. Suh for collaboration on this.}

In \S2 we derive an identity for the black hole S-matrix, under certain plausible assumptions.  This identity determines the leading effect of throwing an additional particle into a black hole that has has formed in the collision of earlier particles.  There is a calculable effect on the outgoing particles that grows exponentially in time, reflecting the Lyapunov behavior, until the calculation breaks down roughly a scrambling time after the extra particle is thrown in.

In \S3 we discuss various puzzles that this result raises.

 \sect{An S-matrix identity}

Refs.~\cite{Dray:1984ha,'tHooft:1990fr,Kiem:1995iy,'tHooft:1996tq} obtained an S-matrix for scattering between a particle falling into a black hole and an outgoing particle very close to the black hole horizon.  The physical significance of this result is puzzling, and it remained somewhat unrecognized until resurfacing recently in the work noted above~\cite{Shenker:2013pqa,Shenker:2013yza,Leichenauer:2014nxa,kitaev,Shenker:2014cwa,Jackson:2014nla,Maldacena:2015waa}.  In fact, the meaning was clearly stated in Ref.~\cite{'tHooft:1990fr}:
\begin{quote}
Suppose the particle falls into a hole that ``planned to emit" a certain series
of particles, possibly to be detected in the late future by some detector. The hole is
then said to be in one of its various possible states in Hilbert space. The infalling
particle, no matter how light it is, will change all that. A different series of particles
will come out. So the incoming particle does cause a transition from one state into a
different one. The effect can be computed rather precisely, using the physics at the
distance scales of the particles considered\ldots
\end{quote}
That is, one obtains a relation between a given black hole S-matrix element and another with an additional ingoing particle.\footnote{This is somewhat in the spirit of a Ward identity or a soft scattering theorem, although we do not see a precise analogy.}
 This is very interesting, and raises further questions.   However,  it was largely overshadowed by a more ambitious attempt in the same work to obtain the full black hole S-matrix, and even to justify the existence of an S-matrix in the first place.  Here we develop the more precise and less speculative interpretation reflected in the quotation above, and discuss its implications.
  
We assume that there is a S-matrix for quantum gravity in $d$ dimensions,
 \be
 \langle 0 | a_{p'_1 \alpha'_1} \ldots a_{p'_m \alpha'_m} a^\dagger_{p_1 \alpha_1} \ldots a^\dagger_{p_n \alpha_n}  | 0 \rangle \,,
\ee
even for processes of such high energy that they are dominated by formation and evaporation of an intermediate black hole.  The $\alpha$ label the internal states of the particles.  A prime will always denote an outgoing particle.  Rather than a momentum basis, it will be important to consider wavepackets $f$ with some localization in time:
 \be
 \langle 0 | a_{f'_1 \alpha'_1} \ldots a_{f'_m \alpha'_m} a^\dagger_{f_1 \alpha_1} \ldots a^\dagger_{f_n \alpha_n}  | 0 \rangle \,.
\ee
We will consider packets interacting at a time when there is a black hole of radius $r_{\rm s}$ centered at the origin, after forming in the earlier collisions.
We will be interested in effects that are large compared to $G r^{2-D}_{\rm s}$, but still less than order one.  For simplicity we suppose that all interactions are irrelevant at low energy, like gravity itself. 

\subsection{From infinity to the horizon}

The important dynamics happens near the horizon, but in order to express the result in terms of the asymptotic S-matrix we must first deal with propagation to the horizon~\cite{Itzhaki:1996rb,'tHooft:1996tq} 
An incoming packet $f$ following a trajectory $r \approx t_0 - t$ partially reflects from the black hole, becoming an outgoing packet $R g' $ at $r \approx t - t_0 + O(r_{\rm s})$ and an ingoing packet $T h $ at tortoise coordinate $r^* \approx t_0 - t$.\footnote{
It is convenient to assume that the added incoming particle is massless, in order that its long-ranged gravitational field not scramble the states of the black hole before the particle itself arrives.}
  We have included reflection and transmission coefficients $R, T$ so that all three packets are normalized in the appropriate inner product  $(f,f)$, e.g.\ the Klein-Gordon inner product for scalars.  The dependence of $R,T$ on $f,\alpha$ is suppressed for convenience.
Defining the corresponding creation operators, $a^\dagger_f = (f,\phi)$ and so on,
we have
\be
a^\dagger_{f,\alpha} = R a^\dagger_{g',\alpha} + T a^\dagger_{h,\alpha}  
\ee
or
\be
a^\dagger_{h,\alpha} = \frac{1}{T} a^\dagger_{f,\alpha} - \frac{R}{T}  a^\dagger_{g',\alpha}  \,. \label{fhg}
\ee

It will be convenient to discuss the addition of a particle in the near-horizon ingoing state $h$, so we see that this is actually a statement about a particular linear combination of asymptotic S-matrix elements with one additional ingoing particle {\it or} one fewer outgoing particle.  
Similarly for outgoing particles,
\bea
a^\dagger_{f',\alpha'} &=& R' a^\dagger_{g,\alpha'} + T' a^\dagger_{h',\alpha'}  \,,
\nonumber\\
a^\dagger_{h',\alpha'} &=& \frac{1}{T' }a^\dagger_{f',\alpha'} - \frac{R'}{T'}  a^\dagger_{g,\alpha'}  \,. \label{fhgp}
\eea

Localized modes $f$ are usually taken to be of positive Schwarzschild frequency.  When expressed in terms of the Kruskal frequencies of the infalling observer, they become a mixture of positive and negative frequencies.  In our analysis, it will be useful for the $f$'s to have compact support in time, and so they must have a mixture of positive and negative frequencies in any time coordinate. To see what is meant by the S-matrix for such modes, consider mixed-frequency modes $b_i$ that are related to purely positive-frequency modes $a_j$ by a Bogoliubov transformation
\be
b_i = A_{ij}  a_j +  B_{ij} a_j^\dagger  \,.
\ee
A number eigenstate for the $b$ modes is equal to a linear combination of number eigenstates for the $a$ modes (start from $N_{\pmb b} = 0$ and work upwards), so an S-matrix element for the $b$ modes is equal to a sum of S-matrix elements with various numbers of $a$ modes.

\subsection{The horizon S-matrix}

The horizon S-matrix~\cite{Itzhaki:1996rb,'tHooft:1996tq,Akhoury:2013bia} is defined in terms of the $h$ basis,
\be
S_{\rm horizon} = \langle 0 | a_{h'_1 \alpha'_1} \ldots a_{h'_m \alpha'_m} a^\dagger_{h_1 \alpha_1} \ldots a^\dagger_{h_n \alpha_n}  | 0 \rangle \,.
\ee
To be precise, the $h$ basis is not meaningful in the early stages when the black hole is forming.  It is probably best to use the $f$ basis over most of the life of the black hole, switching to the $h$ basis during the time of interest.  Alternatively one can use Eqs.~(\ref{fhg},\ref{fhgp}) to express everything in the $f$ basis.   The $h$ basis is well defined during the period when the additional particle is added.

The important effect arises because ingoing and outgoing quanta collide with high energies near the black hole horizon.  Consider an incoming quantum on the trajectory $r^* \sim t_0 - t$ and an outgoing quantum on the trajectory $r^* \sim t - t_1$.  For for $t_1 > t_0$ these meet at $r^* \sim \frac12 (t_0 - t_1)$, at which point their center-of-mass energy in an inertial frame is boosted by a factor
\be
\sqrt{g^{00}} \sim e^{\pi (t_1 - t_0)/\beta} \,,
\ee
where $\beta$ is the inverse Hawking temperature.
Due to this growth, the gravitational interaction becomes important at large $t_1 - t_0$.  

As the incoming particle approaches the horizon, its ultrarelativistic gravitational field takes a simple shock form~\cite{Dray:1984ha},
\be
ds^2 = 2 g_{uv} dv \left(du - F(v, \theta) dv\right) + r^2 d\Omega_{{D-2}}^2 \,.
\ee
Here 
\be
u = - e^{2\pi  (r^* - t)/\beta}\,,\quad v = e^{2\pi  (r^* + t)/\beta}\,, \label{krus}
\ee
 are the Kruskal coordinates, and $\theta$ are coordinates on $S^{D-2}$.  The unperturbed metric, $F(v,\theta)=0$, is Kruskal.  The perturbation satisfies~\cite{Dray:1984ha}
\be
g_{uv}(0,0) \left( -\nabla^2_{D-2} + \frac{(D-3)(D-2) }{2}  \right) F(v, \theta) = -16\pi G \partial_v h(v, \theta) \partial_v h(v, \theta) 
\ee
for incoming packet $h(v,\theta)$.  The ingoing packet and its shock are localized in $v$.   
As the outgoing packet $h'(u, \theta)$ passes through the shock, the leading effect is a shift~\cite{Dray:1984ha},
\bea
\Delta u(\theta) &=& \int dv\, F(v, \theta) \,,
\nonumber\\
 g_{uv}(0,0) \left( -\nabla^2_{D-2} + \frac{(D-3)(D-2) }{2} \right) \Delta u(\theta)  &=& -16\pi G r_{\rm s}^2 \int dv\, \partial_v h(v, \theta) \partial_v h(v, \theta) \,. \quad \label{du}
\eea
The outgoing packet $ \widetilde h'(u, \theta)$ after the shock is related to the packet $h'(u, \theta)$ before by
\be
 \widetilde h'(u, \theta) =  h'(u -  \Delta u(\theta) , \theta)  \,.
\ee
The shift is constant in the smooth $u$ coordinate, and so via the coordinate transformation~(\ref{krus}) one gets
\be
\Delta t(\theta) =\Delta u(\theta) \frac{d t}{d u}  = e^{2\pi  ( t - r^*)/\beta} \Delta u(\theta)\,.
\ee
This grows exponentially with time after the ingoing shock, reflecting the chaotic dynamics of the black hole~\cite{Shenker:2013pqa,Shenker:2013yza,kitaev,Shenker:2014cwa,Maldacena:2015waa}.

This has a simple interpretation.  The infalling particle causes the event horizon to expand.  A mode traveling away from the horizon will find itself somewhat closer to the new horizon, and so take longer to escape.  (In the coordinate system above, the horizon stays fixed at $u=0$ and the particle shifts toward it.)  A mode at radius $r_{\rm s} + \delta$ escapes from the black hole after a time of order 
\be
t \sim \frac{\beta}{2\pi} \ln( r_{\rm s}/\delta) \,.
\ee
Expanding the horizon by $\Delta r_{\rm s}$ effectively reduces $\delta$ by $\Delta r_{\rm s}$, and so
\be
\Delta t \sim \frac{\beta \Delta r_{\rm s}}{2\pi \delta} \sim \Delta r_{\rm s}
e^{2\pi t/\beta} \,.
\ee

We now translate this into a precise statement about the change in the S-matrix.  As an operator statement~\cite{Kiem:1995iy},
\be
a_{\widetilde h',\alpha'} a^\dagger_{h,\alpha} = a^\dagger_{ \widetilde h,\alpha}  a_{ h',\alpha'}   \,. \label{commute}
\ee
The tilde on $a^\dagger_{ \widetilde h,\alpha}$ reflects a similar shift in the ingoing packet, but this packet continues through the horizon so the shift is not seen.
The S-matrix with additional incoming particle $h$ can then be written
\bea
&& \langle 0 | a_{\widetilde h'_1 \alpha'_1} \ldots a_{\widetilde h'_m \alpha'_m} a^\dagger_{h,\alpha}  a^\dagger_{h_1 \alpha_1} \ldots a^\dagger_{h_n \alpha_n}  | 0 \rangle 
\nonumber\\
&&\qquad\qquad\qquad
= \langle 0 | a_{\widetilde  h'_1 \alpha'_1} \ldots a_{ \widetilde h'_{k-1} \alpha'_{k-1}} a^\dagger_{\widetilde  h,\alpha}  a_{  h'_k \alpha'_k} \ldots a_{  h'_m \alpha'_m} a^\dagger_{h_1 \alpha_1} \ldots a^\dagger_{h_n \alpha_n}  | 0 \rangle \,.
\label{oprel}
\eea
That is, we have commuted $a^\dagger_{h,\alpha}$ to the left past outgoing modes $k, \ldots, m$.  In effect, whatever would have been the state of the mode $a_{h'_i \alpha'_i}$ without the additional particle is now the state of the mode $a_{\widetilde h'_i \alpha'_i}$, for $i = k, \ldots, m$.    We commute only as far as we trust the result~(\ref{commute}), a point that we will address below.

This is the naive effect of the expansion of the horizon, but 
the result breaks down beyond the scrambling time.  
For $t_1 - t_0 \gsim (\beta/2\pi) \ln (r_{\rm s}^{D-3}/EG)$, which I will take as defining the scrambling time $t_{\rm scr}$, the shift $\Delta u$ exceeds $-u$ over most of the horizon, and the outgoing particle no longer escapes~\cite{Kiem:1995iy,Itzhaki:1996rb}.  Of course, Hawking radiation does not cease.  Later Hawking modes originate further from the original horizon, and it is an important open question to understand how these become imprinted with information.  Later outgoing particles, as well as the added ingoing particle, undergo scrambling dynamics that is outside the effective field theory approximation.

At  $t_{\rm scr}$, the c.m. energy of the collision is of order
\be
E_{\rm coll} \sim \sqrt{r_{\rm s}^{4-D}/G} \,.
\ee
In $D=4$ this is the Planck energy, but the typical impact parameter is $r_{\rm s}$ and so the gravitational interaction is still weak.  Thus, the cutoff due to the horizon expansion occurs before strong gravity effects such as black hole formation~\cite{Shenker:2014cwa},\footnote{I thank D. Stanford for emphasizing this.} and so we can use the commutator~(\ref{commute}) up to the scrambling time.

If the string scale is rather low, there could still be stringy corrections to the Lyapunov exponent~\cite{Shenker:2014cwa}.  Of course, there are still large nonlocal invariants, and it is interesting to ask whether string effects such as those studied in~\cite{Dodelson:2015toa} could manifest in the S-matrix.

The  relation~(\ref{oprel}) includes the states of particles not appearing in the S-matrix, the outgoers that are captured by the expanding horizon, and the extra ingoer on the right.  In order to write a relation for observables, let us recall the basic intuition: up until the scrambling time, whatever would have been the occupation number for $a_{h'_i \alpha'_i}$ is now the occupation number for $a_{\widetilde h'_i \alpha'_i}$.  Since this should be true in any basis, the full density matrices for these modes should have the same property.  Thus, we consider the square of the S-matrix and trace over all modes coming out after the scrambling time, to get
\bea
&& \sum_{\widetilde X} \langle \widetilde X | a_{ \widetilde h'_k \alpha'_k} \ldots a_{\widetilde h'_m \alpha'_m} a^\dagger_{h,\alpha}  a^\dagger_{h_1 \alpha_1} \ldots a^\dagger_{h_n \alpha_n}  | 0 \rangle
\langle\widetilde X  | a_{\widetilde j'_l \beta'_l} \ldots a_{\widetilde j'_m \beta'_m} a^\dagger_{h,\alpha}  a^\dagger_{j_1 \beta_1} \ldots a^\dagger_{j_n \beta_n}  | 0 \rangle^*
\nonumber\\
&&\quad  = \sum_{  X}  \langle   X |  a_{  h'_k \alpha'_k} \ldots a_{  h'_m \alpha'_m} a^\dagger_{h_1 \alpha_1} \ldots a^\dagger_{h_n \alpha_n}  | 0 \rangle
 \langle   X |  a_{  j'_l \beta'_l} \ldots a_{  j'_m \beta'_m} a^\dagger_{j_1 \beta_1} \ldots a^\dagger_{j_n \beta_n}  | 0 \rangle^* \,.
 \label{result}
\eea
  The notation $ \widetilde X$ reflects the need to shift the time cutoff on the left-hand side to keep the one-to-one correspondence between outgoing modes.  This is a map from the initial state, or density matrix, of the black hole (obtained by contracting the latter with the unprimed indices), to the reduced density matrix for the outgoing modes before the scrambling time.  Thus it is a reduced version of the \$-matrix~\cite{Hawking:1976ra}, but one that preserves unitarity because it is derived from an S-matrix.\footnote{I thank D. Harlow for emphasizing this.}  We claim that this is the precise form of the identity implied by the quotation at the beginning of the section, capturing effects larger than $G r^{2-D}_{\rm s}$.  

It is useful to use wavepackets $h'$ that have finite support in time, so that one can sharply identify those that are before the cutoff.  Since the shift~(\ref{du}) depends on angle and becomes large near the incoming particle, it is also useful to use packets that have limited angular support, avoiding the incoming particle.  It should be noted that the identities~(\ref{oprel},\,\ref{result}) make sense as they stand only for an incoming packet $h$ having narrow support in angle, since the shift depends on this; the packet should not be so narrow that the transverse momentum competes with the large radial momentum.  To go to a general basis, one would first extend to off-diagonal elements in $h$ (which would mean different shifts on the $h'$ and $j'$ packets), and then transform to a new basis.  This would be complicated due to the dependence of the shifts on angle.

 If the incoming particle is thrown in during the first half of the life of the black hole, a typical initial state will map to a maximally mixed density matrix~\cite{Page:1993wv} and there will be no observable effect from the perturbation.  For special initial states, or during the second half of the life of the black hole, the effect will be detectable.  These points are essentially those of~\cite{Hayden:2007cs,Avery:2012tf}.

\sect{Puzzles}

The identity~(\ref{result}) seems compelling, but in a sense it is quite surprising.  The basic input is the near horizon scattering amplitude between ingoing and outgoing particles.  But in a natural black hole there should be no outgoing particles near the horizon from which to scatter!  An infalling particle, or observer, should fall through the horizon smoothly.

Indeed, within a low energy effective field theory description of the horizon, where the Hawking calculation is valid~\cite{Hawking:1976ra}, there is no relation analogous to Eq.~(\ref{result}).  The vacuum state in the neighborhood of the horizon is invariant under small changes of $u$.  The modes exterior to the black hole can only be described by a density matrix, they are not tagged by their entanglement with earlier Hawking modes, and so there is no way to see that something has happened.  This argument also implies that the average Hawking flux is not (substantially) changed by the infalling particle, in spite of the effect on specific packets.

The scattering becomes visible due to the additional assumption that there is an S-matrix.  The argument is a hybrid: we assume that low energy effective field theory breaks down in a manner that allows an S-matrix, but that we can use effective field theory to analyze scattering involving small numbers of quanta outside the black hole (the derivation of the S-matrix involves applying the commutator~(\ref{commute}) to $O(t_{\rm scr}/r_{\rm s})$ modes).  These assumptions are widely employed in applications of \mbox{AdS/CFT}.  We believe that the result~(\ref{result}) has the ring of truth, in connecting the thermal properties of black holes to chaos, and it seems to fit with properties of strongly coupled field theories~\cite{kitaev,Maldacena:2015waa}.   However, it still remains to justify the assumptions from a more fundamental understanding. 

The existence of the identity~(\ref{result}) reflects the fact that the outgoing modes carry information even near the horizon.  In the words of Mathur~\cite{Mathur:2009hf}, the horizon cannot be information-free if there is to be a black hole S-matrix.  
We can illustrate this with an analogous laboratory experiment.  Consider a detector measuring the vacuum fluctuations of a mode that passes through.  What is to be measured is the analog of $N_{ h',\alpha'} = a^\dagger_{ h',\alpha'}    a_{ h',\alpha'}$.  The mode $h'$ has positive Schwarzschild frequency and so mixed positive and negative frequencies as seen by a local inertial observer; its distribution in vacuum is thermal.  The measurement correlates this mode with a register in the detector.  A later observer measuring these same fluctuations can correlate her own detector with the first.  The mode therefore carries information.

But one cannot send information with vacuum fluctuations.  Once the first measurement has been made, the field is no longer in vacuum.  The mode being measured still has a thermal distribution, but it is entangled with the first detector and not with the nearby modes of the field.  Thus, we can be quite certain that if we measure the flux of energy in the neighborhood of the mode after the measurement it will be positive.

The black hole is carrying out a similar measurement on all of the outgoing modes --- this is the meaning of (\ref{result}).\footnote{To be precise, the black hole's measurement is a bit more disruptive.  The lab experiment leaves a correlation between the message modes and their partners, so the message and the register are not in a pure state.  In the black hole the early and late radiation together are in a pure state.  I thank R. Bousso for emphasizing this.}  The associated energy flux is the firewall.
  This is not a new firewall argument but a restatement of the original one~\cite{Almheiri:2012rt}.  The key assumptions  are the same --- an S-matrix, and effective field theory outside the horizon.  But the identity~(\ref{result}) and the laboratory analog make the issue more visceral.  An infalling observer will pass through a beam of information carried by high frequency quanta, and actually affects the passing information.  How then can they not also encounter a beam of energy?\footnote{It was not the intention of the author to make this a paper about firewalls, but all roads seem to lead to them.}

Most attempts to avoid the firewall identify the register in the first detector with the modes of the field near the beam~\cite{AisRB}.  Clearly this does not make sense in the laboratory example, and it is far from clear that it makes sense for the black hole~\cite{AisnotRB}.  Another idea would modify effective field theory outside the horizon~\cite{Giddings:2012gc}, so that~(\ref{result}) would no longer hold.  Finally, we note that~\cite{Hawking:2014tga} has suggested that chaotic behavior of the black hole might be an alternative to the firewall, but we have found that they are two sides of the same coin (or horizon).


\section*{Acknowledgments}

I thank Tom Banks, Raphael Bousso, Steve Giddings, Daniel Harlow, Sunny Itzhaki, Ted Jacobson, Alexei Kitaev, Don Marolf, Ben Michel, Kyriakos Papadodimas, Andrea Puhm, Suvrat Raju, Vladimir Rosenhaus, Steve Shenker, Eva Silverstein, Mark Srednicki, Josephine Suh, and Gabriele Veneziano for helpful dicussions.  This work was supported in part by NSF Grants
PHY11-25915 and PHY13-16748.


\end{document}